\documentstyle[epsfig]{cupconf}

\ifoldfss
\else
  \ifnfssone
    \newmathalphabet{\mathit}
      \addtoversion{normal}{\mathit}{cmr}{m}{it}
      \addtoversion{bold}{\mathit}{cmr}{bx}{it}
    \newmathalphabet{\mathcal}
      \addtoversion{normal}{\mathcal}{cmsy}{m}{n}
    \else
    \ifnfsstwo
    \fi
  \fi
\fi

\def\umu{\mu} 
\def\hexnumber#1{\ifcase#1 0\or1\or2\or3\or4\or5\or6\or7\or8\or9\or
 A\or B\or C\or D\or E\or F\fi }

\makeatletter
\ifx\CUP@mtlplain@loaded\undefined
\else
\fi
\makeatother
 \makeatletter
 \ifx\CUP@mtlplain@loaded\undefined
   \font\tenbmi=cmmib10 at 10pt
   \font\sevenbmi=cmmib10 at 7pt
   \font\fivebmi=cmmib10 at 5pt

   \newfam\bmifam
   \textfont\bmifam=\tenbmi
   \scriptfont\bmifam=\sevenbmi
   \scriptscriptfont\bmifam=\fivebmi
   
 \fi
 \makeatother
\ifnfsstwo

\fi
\ifnfssone

\fi
\ifoldfss

\fi

\mathchardef\varLambda="0103

\makeatletter
\ifx\CUP@mtlplain@loaded\undefined
\else
\fi
\makeatother
\makeatletter
\ifx\CUP@mtlplain@loaded\undefined
  \font\tenbms=cmbsy10
  \font\sevenbms=cmbsy10 at 7pt
  \font\fivebms=cmbsy10 at 5pt
  \newfam\bmsfam
  \textfont\bmsfam=\tenbms
  \scriptfont\bmsfam=\sevenbms
  \scriptscriptfont\bmsfam=\fivebms

  \edef\bsy@{\hexnumber\bmsfam}
  \mathchardef\bnabla="0\bsy@72
\fi
\makeatother
\title[GRB Prompt Optical Emission]{Prompt Optical Emission from Gamma-ray Bursts}

\author[R. Kehoe {\it et al.\/}]%
{Robert Kehoe$^1$, Carl Akerlof$^1$, Richard Balsano$^2$, Scott
Barthelmy$^{3,4}$, Jeff Bloch$^2$, Paul Butterworth$^{3,5}$, Don
Casperson$^2$, Tom Cline$^3$, Sandra Fletcher$^2$, Fillippo Frontera$^6$,
Galen Gisler$^2$, John Heise$^7$, Jack Hills$^2$, Kevin Hurley$^8$, 
Brian Lee$^1$, Stuart Marshall$^9$, Tim McKay$^1$, Andrew Pawl$^1$, 
Luigi Piro$^{10}$, Bill Priedhorsky$^2$, John Szymanski$^2$ and Jim Wren$^2$}       

\affiliation{$^1$University of Michigan, Ann Arbor, MI 48109, USA\\[\affilskip]
$^2$Los Alamos National Laboratory, Los Alamos, NM 87545, USA\\[\affilskip]
$^3$NASA/Goddard Space Flight Center, Greenbelt, MD 20771, USA\\[\affilskip]
$^4$Universities Space Research Association, Seabrook, MD 20706, USA\\[\affilskip]
$^5$Raytheon Systems, Lanham, MD 20706, USA\\[\affilskip]
$^6$Universit\`{a} degli Studi di Ferrara, Ferrara, Italy\\[\affilskip]
$^7$Space Research Organization, Utrecht, The Netherlands\\[\affilskip]
$^8$Space Sciences Laboratory, University of California, Berkeley, 
CA 94720-7450, USA\\[\affilskip]
$^9$Lawrence Livermore National Laboratory, Livermore, CA 94550, USA\\[\affilskip]
$^{10}$Instituto Astrofisica Spaziale, Rome, Italy}

\begin{document}
\ifnfssone
\else
  \ifnfsstwo
  \else
    \ifoldfss
      \let\mathcal\cal
      \let\mathrm\rm
      \let\mathsf\sf
    \fi
  \fi
\fi

\maketitle

\begin{abstract}

	The Robotic Optical Transient Search Experiment (ROTSE) seeks 
to measure contemporaneous and early afterglow optical emission from 
gamma-ray bursts (GRBs).  The ROTSE-I telescope array has been fully 
automated and responding to burst alerts from the GRB Coordinates 
Network since March 1998, taking prompt optical data for 30 bursts in 
its first year.  We will briefly review observations of GRB990123
which revealed the first detection of an 
optical burst occurring during the gamma-ray emission, reaching 9th 
magnitude at its peak.  In addition, we present here preliminary 
optical results for seven other gamma-ray bursts.  No other optical 
counterparts were seen in this analysis, and the best limiting 
sensitivities are $m_V > 13.0$ at 14.7 seconds after the gamma-ray rise, 
and $m_V > 16.4$ at 62 minutes.  These are the most stringent limits 
obtained for GRB optical counterpart brightness in the first hour 
after the burst.  This analysis suggests that there is not a strong 
correlation between optical flux and gamma-ray emission.

\end{abstract}

\firstsection

\section{Introduction}

\subsection{Gamma-ray Observations}

	Fast, intense bursts of cosmic gamma-rays and energetic X-rays 
were first observed about 30 years ago (\cite{grb73}).  Since that
time, satellite missions have determined several characteristics of
these events.  They are generally very brief but are otherwise
extremely diverse in their gamma-ray temporal variations.  Durations 
range from 0.005 to 100's of seconds, and intensity fluctuations are
as short as 0.3 ms (\cite{shorttime}).  They are often instantaneously 
the brightest gamma-ray source in the sky.  Studies of the over 2000 currently
recorded bursts indicate a thoroughly isotropic distribution with no
detectable concentration towards the galactic plane and no
angular correlations with other astrophysical structures (\cite{angcorrel}). 

\subsection{Counterparts at Other Wavelengths}

	Given the lack of a spatial or temporal pattern, 
it has been extremely difficult to comprehend the physical mechanisms 
from the gamma-ray observations alone.  Since the mid-70s, there 
have been many attempts to detect counterparts at other wavelengths,
but they were unsuccessful until 1997.  The difficulty arose from the 
brevity of bursts, the lack of arc-minute localizations and
theoretical prejudices concerning the burst progenitor.  Currently,
the two main GRB efforts utilize the BATSE detectors
(\cite{batse}) on-board the Compton Gamma-Ray Observatory, and the GRBM
(\cite{feroci97}) and WFC (\cite{jager97}) on the BeppoSAX satellite, 
and they have addressed the observational limitations very differently.  
BATSE's advantage is its near complete coverage of the sky, which
allows observation of about 300 bursts per year, and its unique ability 
to provide rough coordinates very rapidly ($\sim 5$ seconds).  These 
localizations are distributed in the form of triggers over the GRB 
Coordinates Network (GCN) (\cite{gcn}, \cite{bacodine}).  Beppo-SAX, on 
the other hand, is able to deliver positions accurate to a few 
arcminutes in a few hours for about a dozen bursts per year.

	The BeppoSAX positions are accurate enough for follow-up with
conventional, small field-of-view telescopes which have detected 
optical counterparts for twelve GRBs.  As a result, we now know that at 
least some GRBs are at cosmological distance (eg. \cite{metzger97}, 
\cite{kulkarni98}) and, if their emission is isotropic, release a 
significant fraction of $M_{\odot}c^2$.  These GRBs, at least, are 
associated with galaxies (\cite{hogg99}).  Studying this afterglow period of 
a few hours to days after the burst has generally revealed a slow, roughly 
power-law decay of optical emission with time.

	These observations, however, are mute concerning the details 
of the burst itself, and there are several limitations in the current
sample.  In particular, the number of such events is small.  There is 
also a bias which arises from the requirement of an observable X-ray 
counterpart.  In addition, because the BeppoSAX GRBM uses a 1 second 
integration period, no counterparts at any wavelength have been 
identified for the class of short ($\sim 0.1$ second) bursts.  Short 
bursts are more likely to have high energy emission, and they occupy a
region of the hardness-ratio vs. duration space well-separated from
long bursts (\cite{shorthard93}, \cite{shorthard95}).  This may imply 
a different type of progenitor for short bursts.  Lastly, these 
observations occur hours after any gamma-ray emission, so that despite 
a growing understanding of afterglows, the burst origin remains a mystery.

\subsection{Prompt, Unbiased Optical Detections}

	Studying early optical emission in an unbiased way has
several advantages.  First, observations at early times may
elucidate details of shock development, the burst environment
and beaming.  Second, by looking for optical emission unbiased
by selections based on burst duration or fluence, we can probe more
thoroughly their range of behavior.

	While we do not know the actual mechanism by which a GRB
occurs, there is a general picture of the development of the cataclysm
aftermath.  A highly relativistic expanding fireball is created in 
which shells develop within the outflow with a spread in velocities.
Gamma-rays emerge from interactions among these shells 
(\cite{meszaros94}, \cite{paczynski94}).
The chaotic time histories of the gamma-rays favor variability from a
central engine such as in the internal shock models (\cite{fenimore96}).
As the relativistic shell propagates into the interstellar medium, its
deceleration produces a forward shock wave (\cite{meszaros92}, 
\cite{meszaros93}) and possibly a reverse shock.
The afterglow is believed to arise from the forward shock, while
significant early optical emission may arise from the reverse shock.
In this scenario, comparisons of simultaneous optical and 
gamma-ray emission comment on the presence and progress of the
external shocks.  For instance, the relative timing of optical and gamma-ray
emission indicates the Lorentz factors involved (\cite{sari99}) as
well as the process by which the shells responsible for the external
shocks arise.

	In addition, we can learn about the environment of the source
at the time of the burst.  The detectability of optical emission alone 
demonstrates that the local environment of the burst is not opaque to 
optical photons.

	Since the relativistic shell is initially moving at very high
bulk Lorentz factors, $\Gamma > 100$, there will be a beaming angle which
is expected to vary roughly as $1/\Gamma$.  If the optical photons 
arise from a more slowly moving region than the gamma-rays, they will 
be more isotropically emitted.  This can occur, for
instance, in the case of later optical emission from the decelerating 
relativistic shell.  As a result, the correlation between this optical 
emission and the gamma-rays will be weak and we could observe a 
bright optical counterpart to a dim or absent gamma-ray burst 
(\cite{rhoads99}).  In addition, the energy release of a GRB might not be
isotropic, being restricted to a jet of angular width, $\theta$.  This
restriction will dominate the optical emission once $\Gamma \sim
1/\theta$.  After some time, however, there will be lateral spreading of
the jet which will further increase the isotropic distribution of the
later optical light (\cite{rhoads99}).

	Any detection of optical emission from short bursts could
reveal their relation to the longer bursts.  The short bursts may 
have a different redshift distribution than that observed for long
bursts.  The two populations might arise from very different sources, 
such as neutron star mergers and hypernovae.  If so, they should have 
distinct distributions within galaxies, and their local environments 
might be quite different.  The fireball mechanism may be entirely 
different for short bursts.  If so, prompt optical observations 
should help illuminate their properties.

\section{The ROTSE Project}

\subsection{Challenges}

	To probe the nature of GRBs, the ROTSE project seeks to 
detect their prompt optical emission.  In particular, 
we wish to detect optical photons coincident with a burst and observe 
optical afterglows to a few hours afterwards.  Ultimately, we wish to 
do this for a sample of GRBs as free of detection biases as possible.  
We have a secondary mission to obtain and provide a stream of 
arcsecond-level positions for many GRBs for more sensitive follow-up.

	 To do this, several technical challenges must be met.  First, 
the gamma-ray emission lasts a few seconds or less so we need to 
respond to triggers 
for gamma-ray bursts in real-time.  Second, they vary rapidly during 
their gamma-ray emission and might be expected to do the same in the 
optical.  Therefore, we must frequently image GRB 
positions to measure short time-scale variation.  All of this requires 
a fully automated system.  Third, the field-of-view of the instrument must 
match the positional accuracy of the trigger which, for BATSE
coordinates, is 5 to 20 degrees.  While this is far larger than the 
field-of-view of conventional optical telescopes, it is achievable in a 
moderately sensitive ($m_V \sim 15$) configuration.  Finally, for analysis 
purposes, we must be able to distinguish rare signals from a
variety of backgrounds.

	We have developed a compact, flexible design consisting of 
fully automated mini-observatories.  In the rest of Section 2, we 
discuss the technical details of our experiment, Section 3 outlines 
the operation of ROTSE-I, Section 4 reviews the observations of 
GRB990123, Section 5 presents results on several other GRBs, and 
Section 6 presents an interpretation of those results.

\subsection{Mini Observatories}

	The ROTSE telescopes are sited in northern New Mexico inside
enclosures providing for the control and protection of the hardware.
These enclosures possess an automatable enclosure cover ('clamshell').  
In general, they are instrumented with weather sensors to detect
rain, clouds, temperature, humidity, and excessive wind.  In addition,
lightning is a serious hazard at the site, so surge suppressors 
must protect all electrical lines to the outside world.  
Uninterruptible power supplies 
(UPSs) perform some power-line filtering and provide about 10 minutes of 
power to gracefully shut down the system in case of power failure.
Each enclosure's internal network and external connection runs on 100 
Mbps ethernet.  At the moment, the site itself is limited 
to a 10 Mbps connection.

	Within the enclosures, there are computers and a custom
control box for operation and monitoring of the various devices.  The 
control box provides power to the telescope mount, clamshell, and the 
weather monitoring devices, as well as communication to the weather 
devices and clamshell.  One of the keys to performing our experiment 
is the utilization of fast, inexpensive PCs. 
The division of labor among the computers in the enclosures varies, 
but in general there is a main computer on which our data acquisition 
system runs.  This includes monitoring the weather devices,
incoming triggers and system errors, control of the clamshell, and
observation scheduling.  Operation of the mount and data processing
may also occur from this computer.  Each camera is interfaced to an 
auxiliary PC via an ISA card interface.

	Our data storage needs are handled by the the Los Alamos Computing 
Division Mass Store System.  This storage system has a several
Petabyte capacity, and provides crucial random access to our large
data set via a quick and convenient interface.

\subsection{Software Components}

	To achieve prompt response times and maximum livetime, each
instrument must be automated, and it must operate in real-time.  We
chose Linux based on its stability, capability and cost, as well as
the avilability of drivers and other software parts necessary for the
experiment.  Although Linux was not inherently designed for real-time
application, we can tolerate 0.1 second latencies in responsiveness
which is well within the operating system's capabilities.

\begin{figure}
\begin{center}
\vspace{2.0pc}
\epsfig{figure=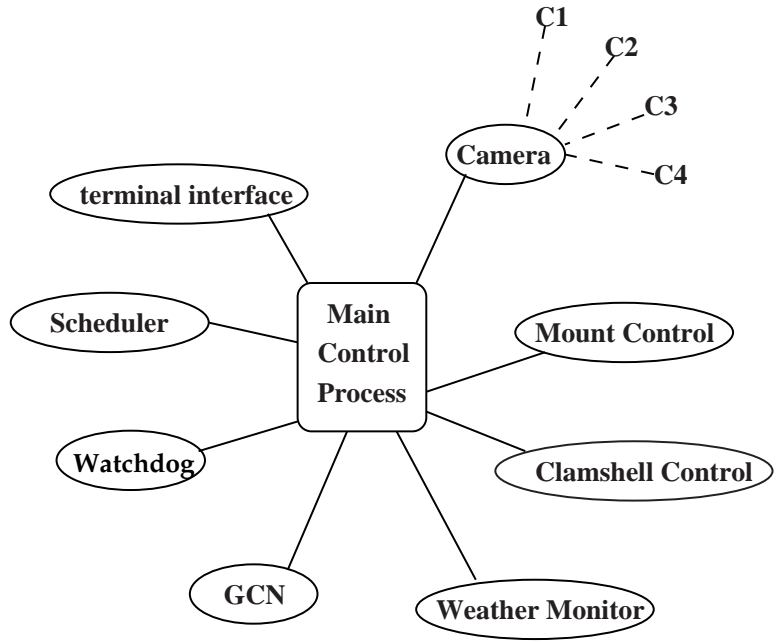,height=3.5in}
  \caption{Scheme of ROTSE data-acquisition system.  A central
	switchboard process running on the control computer channels 
	commands from a user interface and scheduler to hardware 
	control processes via shared data channels (solid lines).  
	The instantaneous status of these processes and other monitors 
	is continuously returned.  Four additional PCs are used to
	control the four CCD cameras, with communication with the main
	computer proceeding over network connections (dashed lines).  
    \label{fig:daq}}
    \end{center}
\end{figure} 

	  We have produced a small suite of programs as diagrammed in 
Figure \ref{fig:daq}.  The overall structure of the system consists of 
a central switchboard process which channels $commands$ from user input 
processes to hardware control programs via shared data structures.  
This switchboard also relays 
$status$ information from the hardware control and monitoring programs 
(ie. camera, mount, clamshell, GCN monitor, weather monitor, and 
watchdog) back to the users.  The two user processes are an astronomical 
scheduler for automation and a modified UNIX shell for manual control.  
Aside from small portions directly interfacing to specific hardware,
we have designed a simple, general structure for easy porting to newer
systems as we develop them.  A large effort has been made to produce a 
responsive system, so we have taken maximum advantage of Linux's 
interrupting and multitasking capabilities.

	In order to be sure of the absolute timing of events, the main
computer is synchronized to public servers using the Network
Time Protocol (NTP), and the camera computers are synchronized
to the main computer also via NTP.  This configuration
maintains the main computer within 10 ms of UTC and the camera
computers within about 1 ms of the main computer and each other.

\subsection{Telescopes}

\begin{figure}
\begin{center}
\vspace{2.0pc}
\epsfig{figure=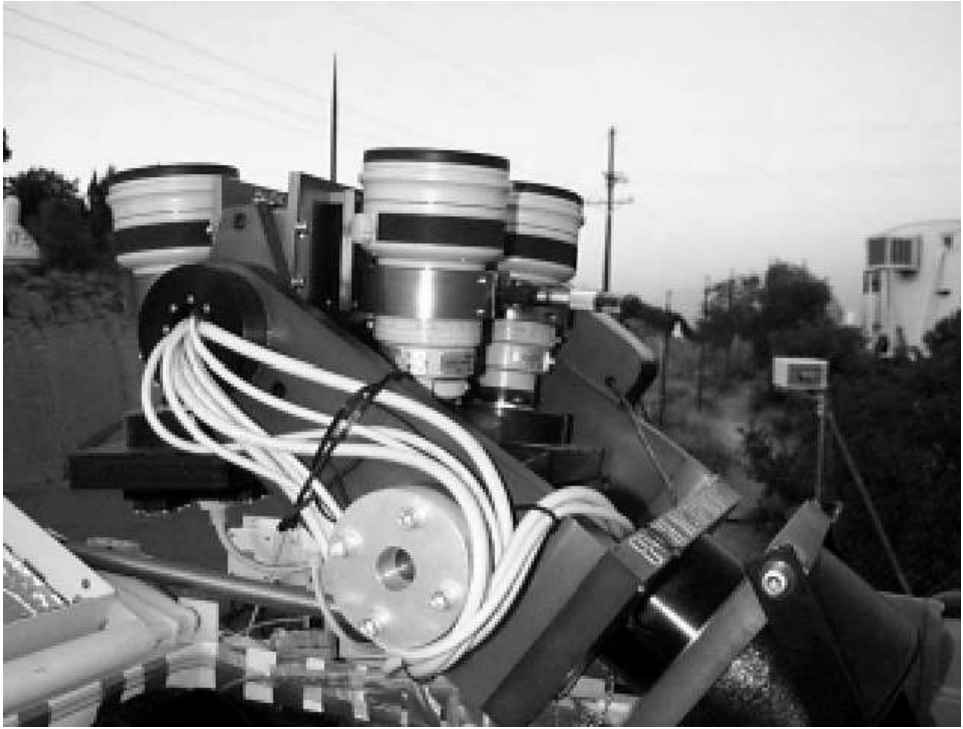,height=3.8in}
  \caption{The ROTSE-I telephoto array.  Four Canon lenses are each
	mounted on CCD cameras on a compact, fast mount.  The
	telescope sits on the roof of the ROTSE-I enclosure.  
    \label{fig:rotsei}}
    \end{center}
\end{figure} 

\begin{figure}
\begin{center}
\vspace{2.0pc}
\epsfig{figure=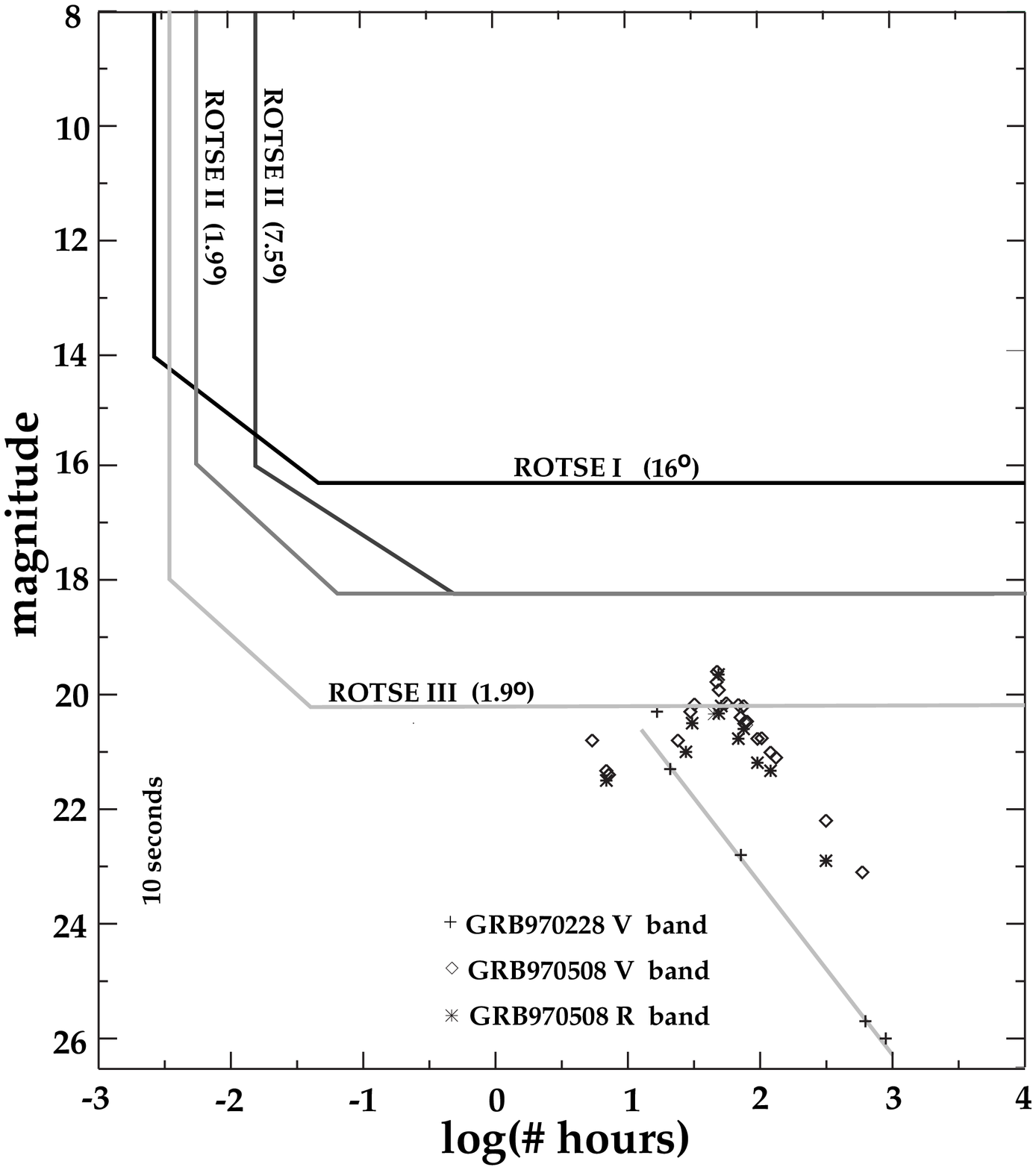,height=5.0in}
  \caption{Sensitivity of ROTSE instruments vs. delay time.  
	ROTSE-I's limiting sensitivity is 14th to
	16th magnitude depending on exposure length, ROTSE-II goes 
	approximately two magnitudes deeper, and ROTSE-III goes four 
	magnitudes deeper.  The field-of-view of each configuration is
	given in parentheses ($7.5^\circ$ for ROTSE-II is with tiling).
	Afterglow points for the first two optical counterparts are 
	shown for comparison (see \cite{galama97}, \cite{fruchter97}, 
	\cite{pedersen98}).
    \label{fig:sensitivity}}
    \end{center}
\end{figure}

	We have developed a two-tiered program to cover the initial
outburst and afterglow periods.  ROTSE-I is fast, wide field-of-view, 
and moderately sensitive.  It consists of a $2 \times 2$ array of
small telescopes co-mounted on a rapid-slewing platform (see Figure 
\ref{fig:rotsei}).  Each telescope is instrumented with a 
thermoelectrically cooled CCD camera employing a 14-bit Thompson 
TH7899M chip with $2048 \times 2048$ 14 micron pixels. Read noise is 
$\sim 25e^-$, and readout is limited by 
the ISA interface to take about 7 seconds.  The optics of each 
telescope consist of a Canon FD 200 mm $f/1.8$ telephoto lens.  We have 
equipped each with a focus-ring clamp 
positioned by a micrometer for accurate manual adjustment.  Our 
sensitivity to faint point sources is maximized by the match of the 
optical point spread function to the pixel size.  The plate scale for 
ROTSE-I images is 14.4"/pixel.  To further improve sensitivity, the cameras 
are operated without filters, and the peak response is in the R, V 
and I bands.  Each telescope is sensitive to 14th magnitude in a 5 
second exposure, and the array covers a $16.4^\circ \times 16.4^\circ$ 
field-of-view.  The mount is capable of slewing to any point in the
sky in less than 3 seconds.  As shown in Figure \ref{fig:sensitivity}, 
ROTSE-I is capable of seeing optical counterparts as dim as 14th 
magnitude by 10 seconds after a burst, and longer exposures achieve 
16th magnitude sensitivity. 

	 The second stage of the experiment brings significant 
improvements in sensitivity to faint objects.  Each of the two existing 
ROTSE-II telescopes consists of a wide-field modified Cassegrain
optical tube instrumented with the same cameras as ROTSE-I and 
a mount with an average slew time of 15 seconds.  We have started building an 
additional set of eight similar telescopes, called ROTSE-III, with improved 
optics, back-illuminated CCD chips, and substantially faster slew times.  
All of these consist of $f/1.9$ optical assemblies with 45 cm apertures 
and $1.9^\circ$ field-of-view.  The plate scale is 3.4"/pixel.  With
these telescopes, we will study optical bursts for those few, prompt, 
accurate localizations from HETE2 (\cite{hete2}), and we will observe optical 
afterglows by scanning the neighborhood of BATSE burst locations.  
We will also search for non-triggered fast-fading optical transients 
which might have a similar physical origin.  The estimated sensitivity
of these instruments is shown in Figure \ref{fig:sensitivity}.

\section{ROTSE-I Operations}

\subsection{Observation Scheduling}

	The astronomical scheduler is responsible for starting and
stopping a night's run, designing observing sequences during
the night, and scheduling darks for image noise correction.  There are 
currently two main observing modes.  Most of the time is spent in a lower 
priority sky patrol.  Given the ROTSE-I field-of-view, 206 frames
cover the celestial sphere with reasonable overlaps.  We observe all fields 
with elevation $> 20^\circ$ in two successive images taken twice nightly.
These images were 25 second exposures until December 1998, and
generally 80 seconds since then.  These data are valuable for 
untriggered transient studies and calibrations.  They also provide 
precursor images for GRB fields 
and permit studies of the optical transient background.

	About once per week, an observable trigger is received via
GCN, and in these instances we interrupt any ongoing sky patrol 
observations for the higher priority alert.  A response is 
scheduled which depends on the trigger coordinates and type.
Different trigger types arrive with different transient position errors
as well as different delays from the initial event detection.  
About half of all triggers received correspond to classic GRBs.
In general, a 
series of exposures with increasing durations is taken as the 
response progresses.  Until December 1998 we employed exposure lengths 
of 5, 25 and 125 seconds, then changed briefly to 5, 75, and 200
seconds, and since January we have used 5, 20, and 80 seconds.  If we 
are observing the burst
within seconds of its rise, we take short exposures initially.  Longer 
delays begin with longer exposures.  If the trigger is of a type with 
a large position error, then we also 'tile' around the given position 
($32^\circ \times 32^\circ$) at specific 
points in the sequence.  In this case, several direct-pointing images 
are taken and then a pair of images is taken in each of the four corners 
around the trigger coordinates.  We then return to the direct 
pointing with longer exposures and begin the sequence again.

\subsection{Online Data Processing}

	Every observing night, multiple raw darks are taken for each
exposure length and averaged to produce a reference dark.  Flats are 
produced by dark-subtracting and median-averaging $\sim 60$ sky patrol 
frames.  For the most part, the flat variation is dominated by 
vignetting which amounts to a 60\% loss at the frame corners.  The 
process of making flats and darks also generates diagnostics which 
are regularly examined for signs of hardware problems.  These 
correction frames are applied to the rest of the data to compensate 
for CCD noise and photometric response variations.  After the 
correction procedure, images are reduced to lists of objects using 
SExtractor (\cite{sextractor}) which provides rough photometry and 
cluster shape information.  Due to processing and data transfer 
limitations, only the triggered data and some of the sky patrol data 
can currently be processed online.  Once a night's observing is done, 
the data is moved automatically to mass storage.

\subsection{Astrometry and Photometry}

\begin{figure} 
\begin{center}
\vspace{2.0pc}
\epsfig{figure=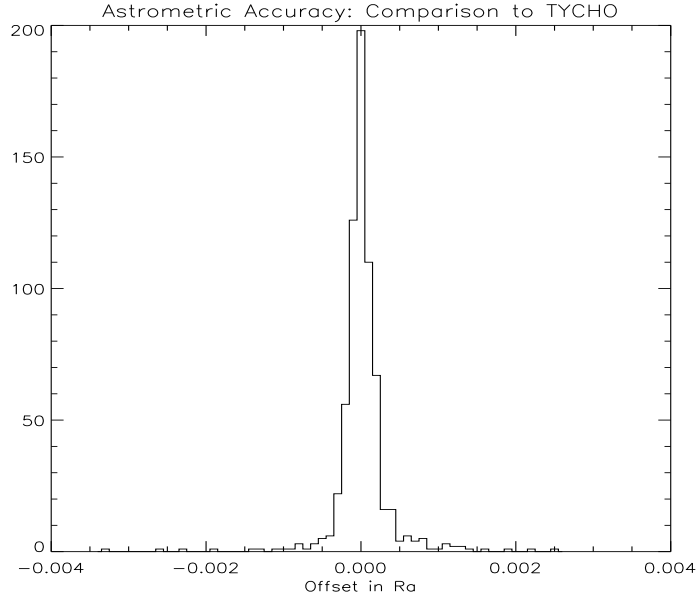,width=4.0in,height=3.2in}
  \caption{RA residuals in degrees for calibrated objects identified in
	ROTSE-I images compared to Hipparcos coordinates.  Centroids
	are accurate to 1.4 arcecond (ie. 0.1 pixel).
    \label{fig:astrometric_accuracy}}
    \end{center}
\end{figure}
\begin{figure}[btp]
\begin{center}
\vspace{2.0pc}
\epsfig{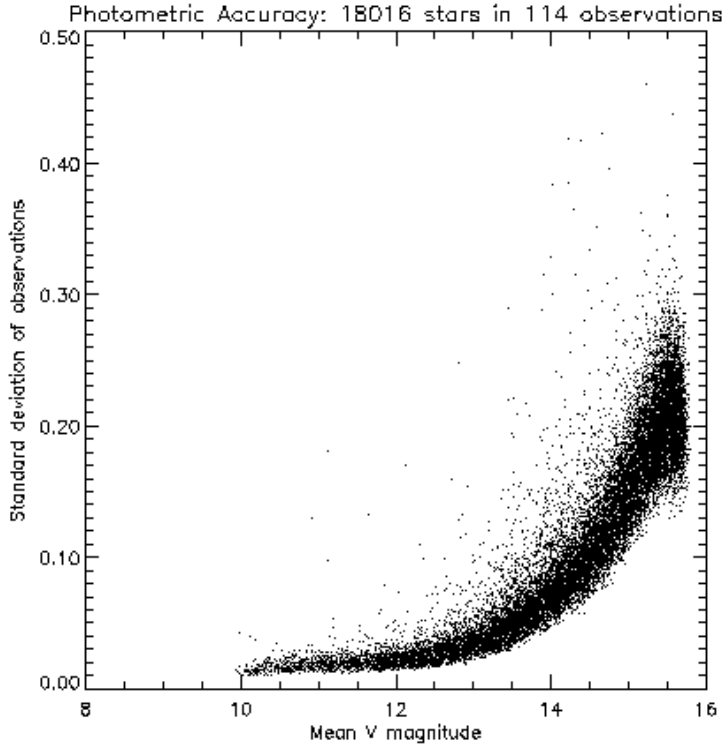}
  \caption{Magnitude residuals for calibrated objects identified in ROTSE-I
	images compared to Hipparcos photometry.  Our final final
	photometric errors are $\sim 0.02$ magnitude for bright stars.
    \label{fig:photometric_accuracy}}
    \end{center}
\end{figure}

	We currently perform our final astrometric and photometric
calibrations offline by comparing our raw object lists to Hipparcos 
data (\cite{tychocat}).  Photometry is established by comparing raw ROTSE 
magnitudes to V-band 
measures and color correcting based on B-V.  Astrometry is determined
by triangle-matching approximately 1000 catalog stars to each image,
and determining warp corrections via a third order polynomial fit.
As shown in Figure \ref{fig:astrometric_accuracy}, our astrometric 
errors are 1.4 arcsec, and Figure \ref{fig:photometric_accuracy} 
indicates our photometric errors to be as good as 0.02 magnitude for
bright stars.

\subsection{Run Summary}

\begin{figure}
\begin{center}
\vspace{2.0pc}
\epsfig{figure=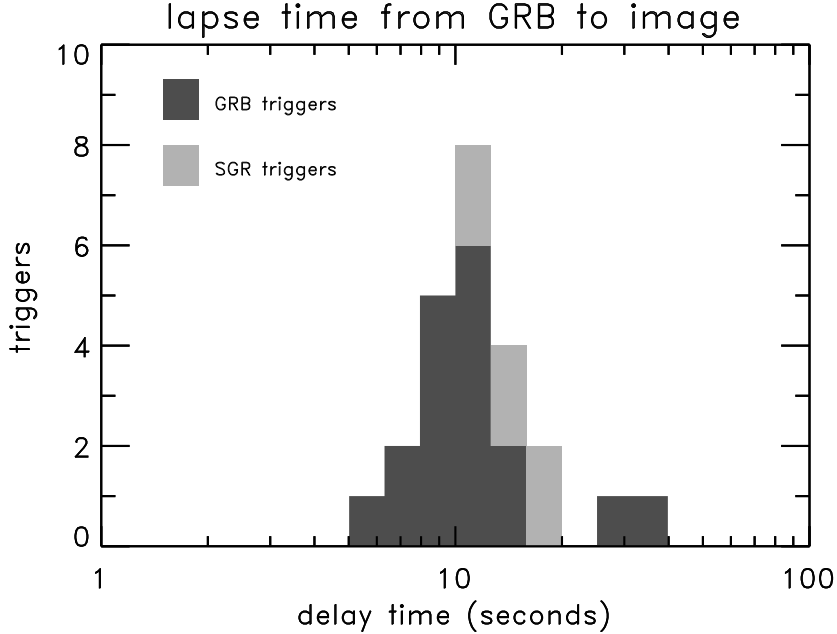,height=3.5in}
  \caption{Delays for prompt GRB triggers.  Times refer to period
	between the gamma-ray rise and the opening of the shutter for the
	first ROTSE-I exposure.  The typical delay time is 10 seconds.
    \label{fig:rotse_delay}}
    \end{center}
\end{figure}

	ROTSE-I operations achieved robust automation in March 1998, and
during the first 12 months observed approximately 75\% of all nights.
The downtime resulted from very bad weather and from occasional
hardware and software failures.  In a typical night, the entire visible 
sky is imaged to 15th magnitude four times.  In the first year, every 
field north of declination $-30^\circ$ was observed between 200 and
900 times.  The data stream is approximately 8 Gb/day, and the total 
amount of data generated is currently $>$ 2 Tb.  In that time, ROTSE-I 
responded to 49 physically interesting triggers.  Of these, 30 were
from classic gamma-ray bursts, 13 were from soft gamma-ray repeater 
events (10 of SGR1900+14), and 6 were X-ray transients.  Response 
times for the subsample of prompt GRB and SGR triggers are shown in Figure
\ref{fig:rotse_delay}.

\section{Contemporaneous Optical Emission from GRB990123}

\begin{figure}
\begin{center}
\vspace{2.0pc}
\epsfig{figure=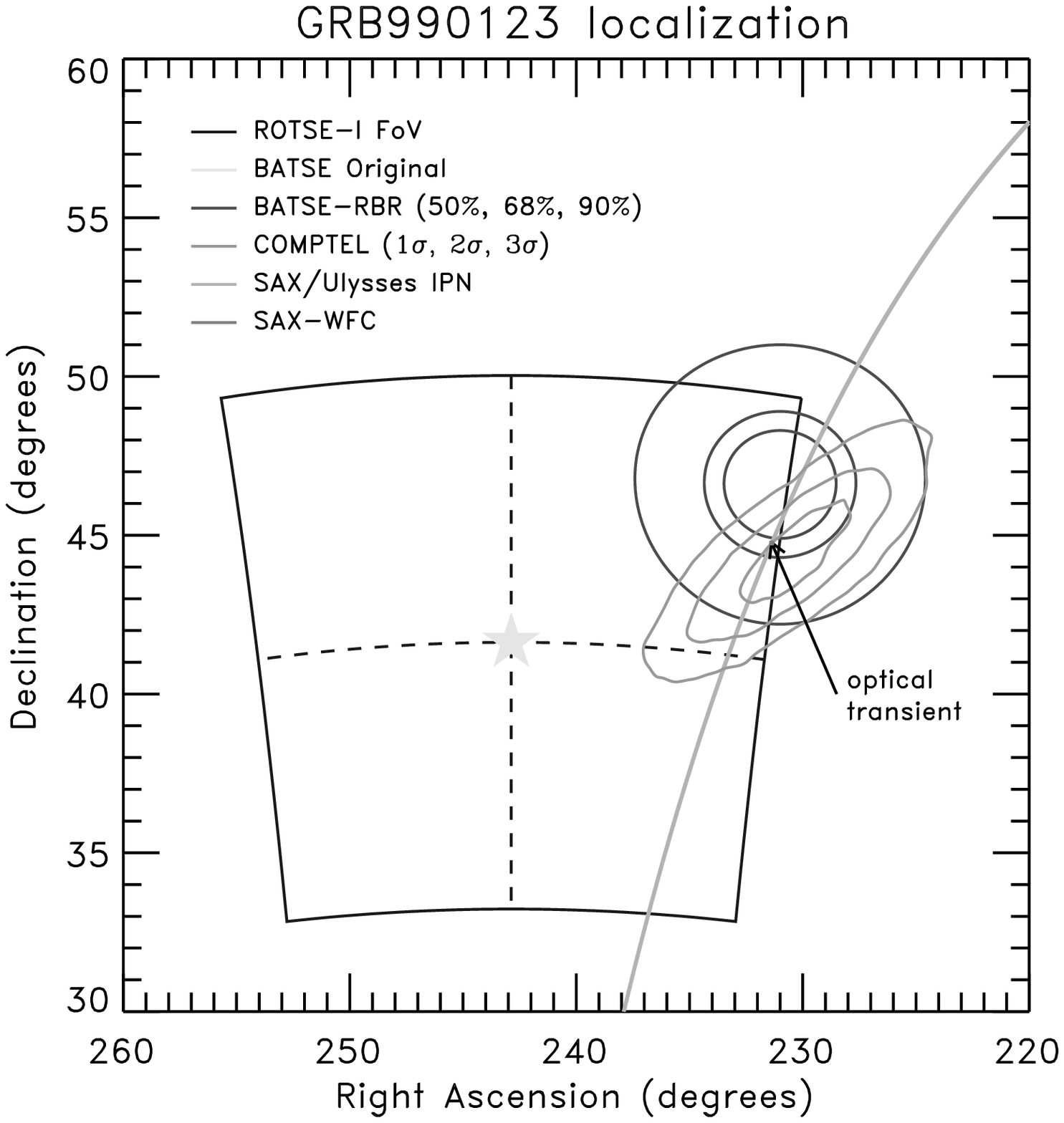,height=4.0in}
  \caption{Various localizations of GRB990123 superimposed on ROTSE-I 
     	field-of-view.  The optical counterpart was within $0.1^\circ$ of
     	the frame edge.
    \label{fig:loc990123}}
    \end{center}
\end{figure}
\begin{figure}
\begin{center}
\vspace{0.0pc}
\hspace{-1.0in}
\epsfig{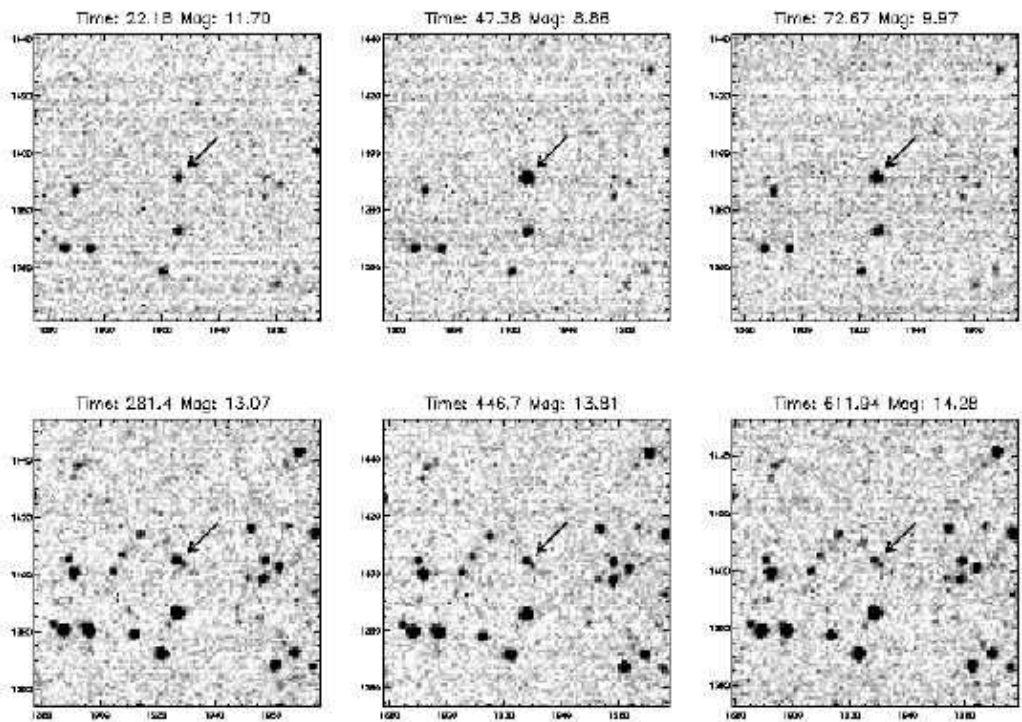}
\vspace{-1.0in}
  \caption{$100 \times 100$ pixel subimage (24 arcminutes across)
	surrounding GRB990123 optical counterpart.  The top row shows 5 
	second exposures, while the bottom row shows 75 second exposures.
    \label{fig:7343_im}}
    \end{center}
\end{figure} 
\begin{figure}[btp]
\begin{center}
\vspace{2.0pc}
\epsfig{figure=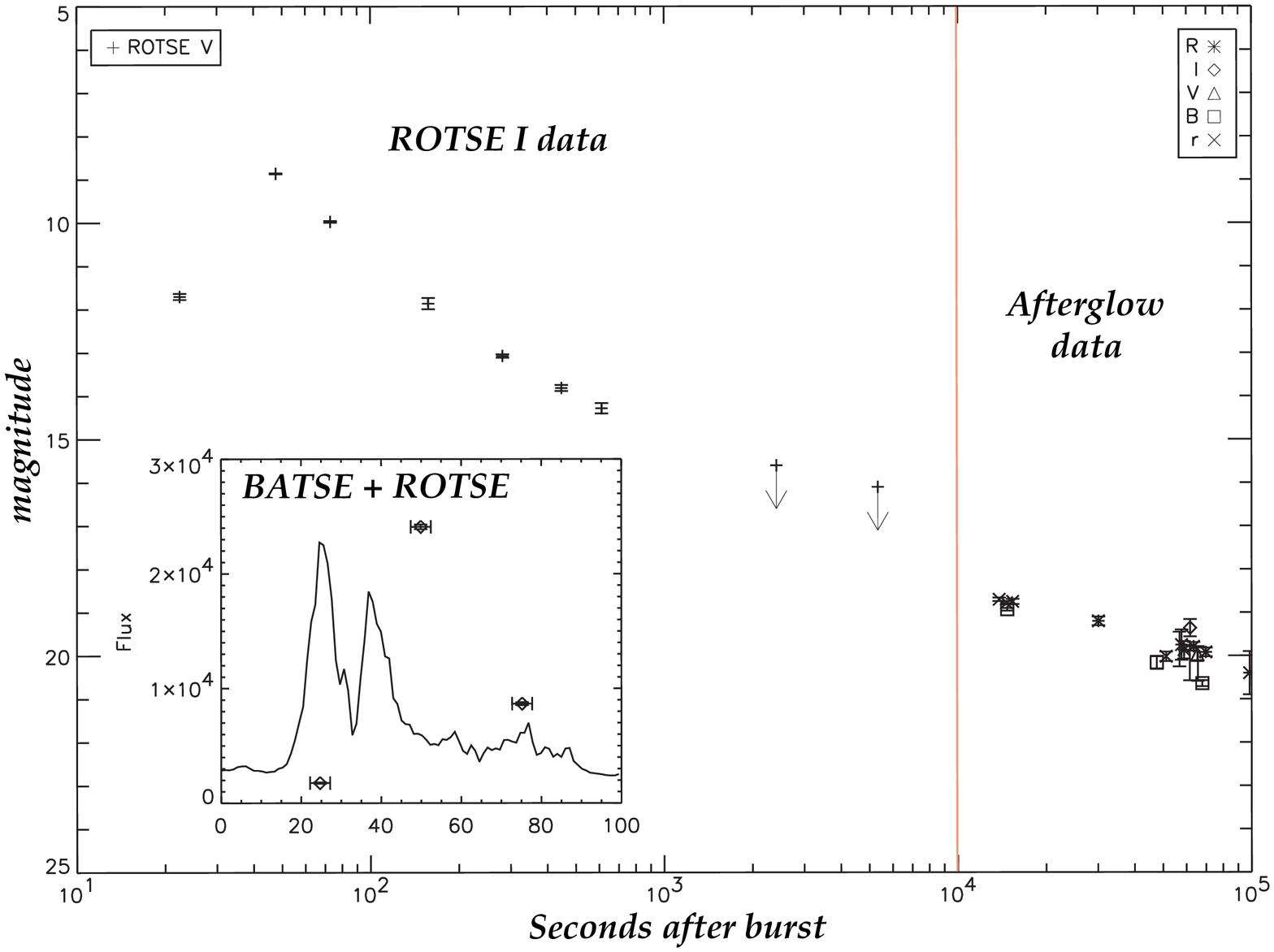,height=3.5in}
  \caption{Lightcurve from ROTSE-I observations with some early
	afterglow points.  The inset superimposes the first three
	ROTSE points on the BATSE gamma-ray lightcurve.  The vertical
	line dividing the ROTSE data from the afterglow observations 
	indicates approximately when the earliest arcminute positions
	can be made available. 
    \label{fig:7343_lc}}
    \end{center}
\end{figure}

	At approximately 9:47 UTC on Jan 23, 1999, BATSE and Beppo-SAX
triggered on an intense GRB.  After 4 seconds, ROTSE received the 
estimated GCN coordinates (see Figure \ref{fig:loc990123}), and
scheduled an observational sequence which began the first exposure after
another 6 seconds.  One hundred minutes and 200 images later, the 
ROTSE trigger response was complete.  In about 4 hours, the Beppo-SAX 
position, accurate to $0.1^\circ$, became available (\cite{gcnc199}), 
leading to the detection of the optical afterglow (\cite{ode201}). 
It also permitted quick identification of a burst of optical photons
in the ROTSE images spatially coincident with the optical afterglow
(see Figure \ref{fig:7343_im}).  As shown in Figure \ref{fig:7343_lc},
this 'optical burst' was surprisingly bright, reaching 9th magnitude
at its peak 50 seconds after the gamma-ray rise (\cite{rotse99}).

	The observations of GRB990123 demonstrate several things about
gamma-ray bursts.  First, some subset of GRBs do exhibit optical
bursts with fluctuations as violent as the gamma-ray variations (see 
inset in Figure \ref{fig:7343_lc}).  This observation and the optical 
brightness of the event imply that optically absorbing material local 
to the event was minimal.

\begin{sloppypar}
	The fact that the optical and gamma-ray emission were both
intense and similarly short-lived suggests that the two are connected.
On the other hand, while the gaps between the optical observations do 
not permit an exact location of the optical peak, it clearly did not 
occur when the gamma-rays were at maximum, thereby suggesting the 
gamma-rays and optical emission came from different processes in the 
burst.  In particular, the observations are consistent with coming
from a reverse shock (\cite{sari99}).
\end{sloppypar}

	When considering the measured redshift, $z = 1.6$, of this GRB
(\cite{iauc7096}, \cite{gcnc219}), it becomes evident that this is a 
truly collossal event.  In optical light, it is the most luminous
object ever recorded, having a peak absolute magnitude of -36.4.  Assuming
isotropy, over $M_{\odot}c^2$ was released in gamma-rays.  Such an
energy output is large enough to cause great difficulty
for most GRB models, which typically provide only about 1\% of the 
inferred energy (see \cite{janka96}, \cite{ruffert97} and
\cite{meszaros97} and references within).  This has led to speculation 
that the emission is not isotropic, suggesting a beaming scenario 
(\cite{kulkarni_nature}).

\section{What About Other Bursts?}

\subsection{Strategies}

	The gamma-ray fluence of GRB990123 is about 100 times that of 
a median BATSE burst.  If there is a strong linear correlation between 
gamma-ray fluence and optical flux, ROTSE-I should be sensitive enough 
to find optical counterparts to roughly half of the GRBs observed.  Our 
ongoing analysis of seven earlier bursts (see Table \ref{tab:sevengrbs}) 
might then be expected to reveal more optical counterparts.
\begin{table} 
  \begin{center} 
  \begin{tabular}{cccccc}
	date & trigger \# & loc. source & coverage (\%) & dur. (sec.)
  & rel. fluence (\%)\\
	\hline
	980329a	& 6665 & SAX & 100 & 55 & 32\\
	980401	& 6672 & IPN & 100 & 37 & 8\\
	980420  & 6694 & IPN & 85 - 100 & 40 & 8\\
	980527  & 6788 & BATSE & 86 & 0.1 & 1\\
	980627  & 6880 & IPN & 60 & 14 & 1\\
	981121  & 7219 & IPN & 68 - 100 & 60 & 7\\
	981223  & 7277 & IPN & 100 & 60 & 13\\
  \end{tabular}
  \caption{Characteristics of seven bursts responded to by ROTSE-I.
  	The second column gives the BATSE trigger number.  The third
	column specificies the origin of the best localization.  The fourth
  	column indicates the coverage of the GRB probability in percent.
 	The fifth column indicates the duration in seconds.  The last 
	column indicates the fluence of the burst as a percent of GRB990123.
	\label{tab:sevengrbs}}
  \end{center}
\end{table}

	We have simplified the analysis by choosing those with the
smallest position errors -- six of the seven possess square-degree 
level localizations or better, while one (GRB980527) has a
BATSE statistical error of about $1.1^\circ$.  This results in an
enormous reduction in background ($> 200\times$).  Aside from the 
one BeppoSAX position for GRB980329a, the more accurate
positions arise from the use of the gamma-ray detectors on-board 
Ulysses (\cite{ulysses}).  By comparing timing information 
between two widely spaced detectors, thin 
annuli are generated which are several degrees long but only about 
$0.1^\circ$ wide (\cite{ipn}).  The intersection of the BATSE position 
probability distribution with such an 'Interplanetary Network' (IPN) 
timing annulus produces an IPN arc.  The downside to restricting
ourselves to using these localizations is that they cannot be 
obtained for GRBs on the faint end of the BATSE fluence distribution.

	We are currently using several analysis strategies to check
the consistency of our methods.  The simplest is a lightcurve
analysis which looks for sources varying by $> 2$ magnitudes in a 
trigger response.  We also match our object lists to more complete 
catalogs such as USNO (\cite{usnocat}) to identify any new objects.  
The most sensitive method we employ is image differencing.  
A template for our trigger response images is constructed from precursor
sky patrol images and subtracted from the triggered data.  
If we do not have a very precise localization such as from an X-ray, 
optical, or radio counterpart, new or varying objects are only 
considered bona fide optical counterpart candidates if they appear in 
at least two successive images.  This is to remove the background 
arising from ghosts, cosmic rays, satellite glints, etc. which show up in 
nearly every image with the field-of-view of ROTSE-I.
If we have an IPN arc, we search 
through the unconstrained IPN annulus in these images for interesting 
objects.  If no source is found, limits are obtained whenever coverage exceeds 
50\% of the IPN arc.  Our results take the form of average magnitudes, 
$\langle m_V \rangle_x$, during an exposure length, $x$, vs. the time, 
$t_+$, after the start of the burst.  Limits refer to the faintest 
$\langle m_V \rangle_x$ to which we are $>$ 50\% efficient at finding 
objects after our analysis selection.  Unless specifically noted, they 
do not correspond to long integration times (ie. hundreds of seconds) 
obtained from co-adding multiple images.  The results discussed here 
are preliminary, and a final, more robust analysis will exploit the 
full sensitivity of the telescope.

\subsection{GRB980329a}

\begin{figure}
\begin{center}
\vspace{2.0pc}
\epsfig{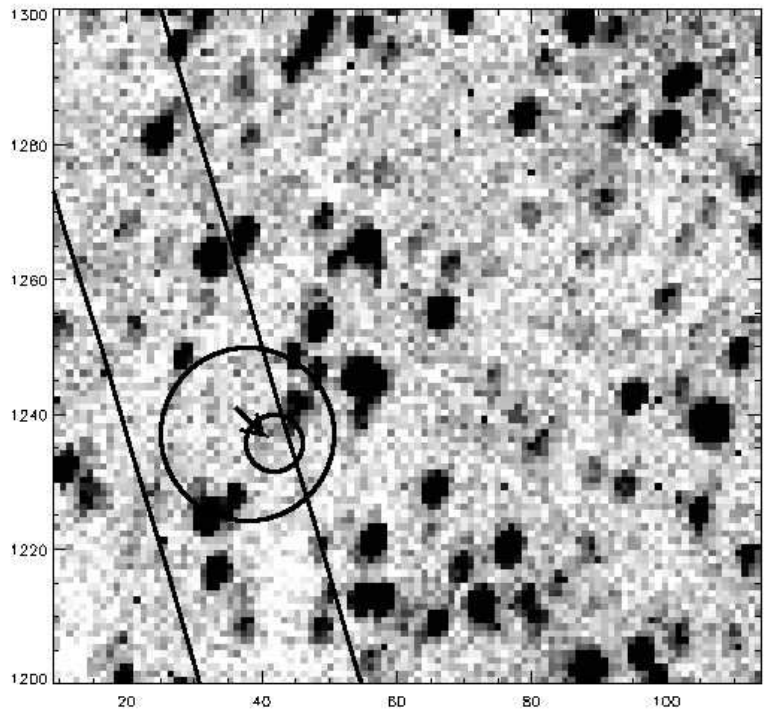}
  \caption{Subimage surrounding GRB980329a radio counterpart.  Three
	images were co-added, giving an effective exposure length of 
	375 seconds.  Circles denote the BeppoSAX localizations
	(\cite{329a_xray}).
    \label{fig:6665_circles}}
    \end{center}
\end{figure} 

	On March 29, 1998, ROTSE-I successfully responded to its
first trigger in automation.  The first exposure was begun 11.5
seconds after the gamma-ray rise, and the complete response was finished
approximately one hour later.  Unfortunately, the sky was fairly
cloudy until the last three frames were taken, and these are 
somewhat hazy.  There were, however, optical (\cite{gcnc41}, 
\cite{gcnc48}, \cite{gcnc52}), X-ray 
(\cite{329a_xray}) and radio (\cite{329a_radio}) counterparts 
observed hours later with the result that the
burst location is known precisely.  Despite the poor quality of the
early data, some images are clear enough in the immediate region of 
the burst to see any 9 - 11 magnitude objects.  Since we know exactly 
where the source is in this case, we use the following reference stars 
in to estimate the sensitivity of the image in the region: 
SAO 59687 ($\alpha = 105.297$, $\delta = 39.177$, $m_V = 8.24$), 
SAO 59692 ($\alpha = 105.308$, $\delta = 38.859$, $m_V = 9.63$), 
SAO 59708 ($\alpha = 105.585$, $\delta = 39.169$, $m_V = 8.65$) 
and GSC 958 ($\alpha = 105.583$, $\delta = 38.883$, $m_V = 10.09$).
To be conservative, we take the sensitivity to be equal
to the magnitude of the dimmest of these objects that can be reliably 
observed in a given image.  The last three images were co-added to 
produce an image (see Figure \ref{fig:6665_circles}) sensitive to 
about 14.8 magnitude as obtained from comparison with the USNO catalog.
No object is detected at the known source location in our images.  The
limits are summarized in Table \ref{tab:sevenlims}.

\subsection{GRB980401}

\begin{figure}
\begin{center}
\vspace{2.0pc}
\epsfig{figure=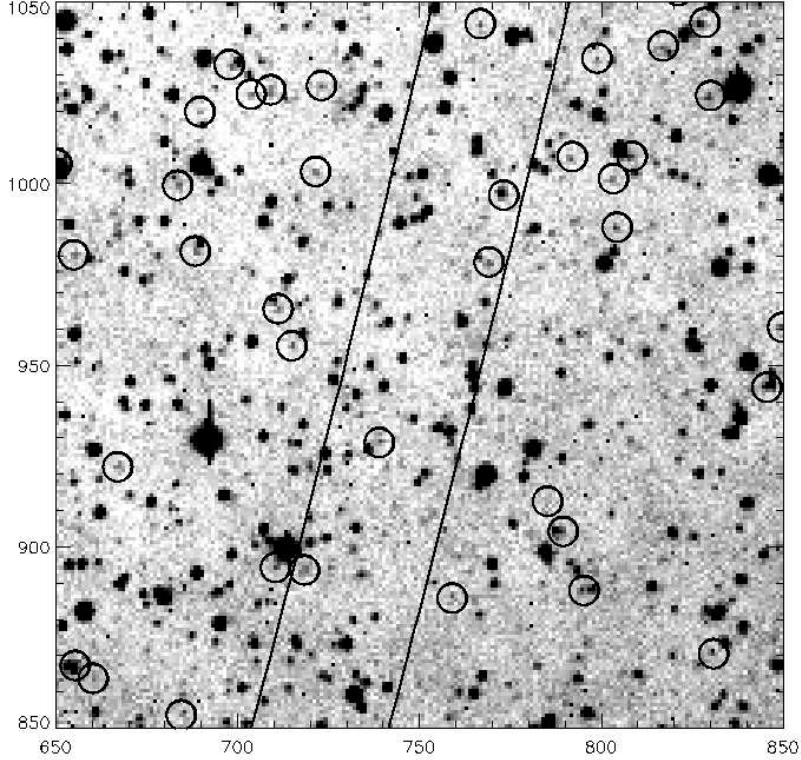,height=4.0in}
  \caption{Subimage from GRB980401 data with portion of overlapping 
	IPN arc (delineated with solid lines).  Circles indicate catalog
	objects with magnitudes between 16.0 and 16.4.
    \label{fig:grb980401}}
    \end{center}
\end{figure} 

     Our third trigger, GRB980401, arrived during focusing tests, so a
manual response was performed starting 2477 seconds after we received the
trigger.  This burst has an IPN localization which is completely 
contained in one camera.
The last four 125 second exposures were co-added into one exposure.
While the total exposed time is 500 seconds, the manual nature of the
trigger spreads the integration time over about 900 seconds.
The resultant image is sensitive to
$\langle m_V \rangle_{500} = 16.4$ at $t_+$ = 62 minutes and shows no 
unkown objects when compared to the USNO catalog (see Figure 
\ref{fig:grb980401}).  This is the most
stringent optical upper limit on a GRB in the first hour timeframe.

\subsection{GRB980420}

	Although conditions for GRB980420 were very good, the camera 
containing most of the final IPN arc in the initial images was out of focus,
thereby greatly reducing the image sensitivity.  In tiled and later 
response images, in-focus cameras overlap the IPN arc resulting in a
greater sensitivity.  No optical source was observed which varied by
more than 2 magnitudes, with an analysis sensitive to 14th magnitude
objects in the later images.

\subsection{GRB980527}

	While much more difficult, we have begun analyzing bursts for
which only BATSE positions are available.  Such an analysis requires 
image differencing to work efficiently.  One such burst is GRB980527.  
The ROTSE-I array covers 100\% of the statistical and 86\% of the 
combined statistical and systematic error region, where we have used 
the systematic error parametrization found in \cite{batse_error}.
The first observation starts at $t_+$ = 12.6 seconds.  Our preliminary 
analysis is sensitive to $\langle m_V \rangle_5 = 13.0$ and 
$\langle m_V \rangle_{25} = 14.2$.  No candidates were found in the 
early non-tiled images.  It should be remembered that this is the only 
short burst in this sample.  The limit of 13.0 at 15 seconds is the 
best limit obtained by any experiment so soon after a burst.

\subsection{GRB980627}

	GRB980627 was fairly dim in gamma-rays and the IPN arc was located
very far from the original trigger localization.  As a result, a 
majority of the probability distribution is only covered in our 
tiled images.  No objects were found which varied by more than 2
magnitudes in this data.  The images are sensitive to approximately
magnitude 12.5, irrespective of exposure length.

\subsection{GRB981121}

	We responded to a GRB on November 21, 1998 which was fairly intense in 
gamma-rays.  The data were taken under good conditions, although our 
shutters were occasionally not opening completely due to very cold 
weather.  Despite this problem, we cover most of the IPN arc for this
burst.  No unidentified objects were seen with a sensitivity to 
$\langle m_V \rangle_5 = 12.8$ in early exposures and 
$\langle m_V \rangle_{25} = 14.3$ in later exposures.

\subsection{GRB981223}

	GRB981223 was another burst which was bright in gamma-rays.
Our trigger response was prompt and the weather was clear.  No 
unidentified objects were seen with limits in the range 12.4 to 13.5.

\section{Results}

\begin{figure}
\begin{center}
\vspace{2.0pc}
\epsfig{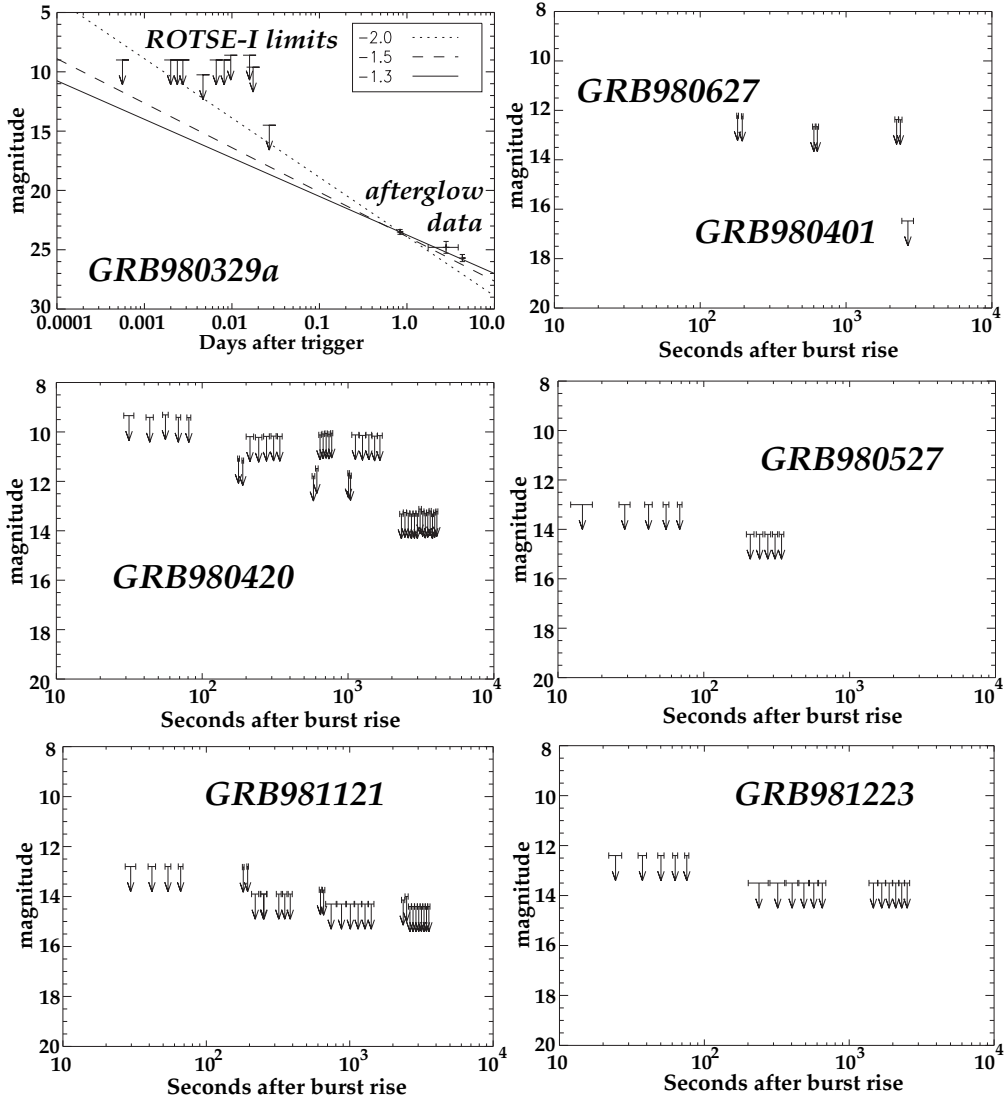}
  \caption{Magnitude limits for seven GRB data sets.  Each plot shows
	apparent V band magnitude vs. time after the gamma-ray rise.  
	The time history for GRB980329a has three R-band afterglow 
	observations (\cite{gcnc41}, \cite{gcnc48}, \cite{gcnc52}) 
	superimposed, along with three power law decays passing
	through the earliest optical detection.
    \label{fig:grb_timehist}}
    \end{center}
\end{figure}

	The preliminary limits placed on the early optical emission 
for these seven bursts are shown in Figure \ref{fig:grb_timehist}, 
and a summary of the limit results is given in Table 
\ref{tab:sevenlims}.  We are able to place a constraint on the 
overall power-law decline of optical emission from GRB980329a to be 
shallower than -2.0 with respect to the afterglow points.
The best limits are currently $\langle m_V \rangle_5 > 13.0$ at 14.7 
seconds for GRB980527, and $\langle m_V \rangle_{500} > 16.4$ at 62 
minutes for GRB980401.  Given the measurements presented, we can
conclude that bright optical counterparts (ie. $m_V \sim 10$) are
uncommon.

\begin{table} 
  \begin{center} 
  \begin{tabular}{cccccccccc}
	date & $t_1$ (s) & $t_{exp1}$ & $m_V(t_1)$ & 
	       $t_2$ (s) & $t_{exp2}$ & $m_V(t_2)$ & 
	       $t_3$ (s) & $t_{exp3}$ & $m_V(t_3)$\\
	\hline
	980329a	& 51.0 &   5 &  9.0 & 415 & 25 & 10.1 & 2231 & 375 & 14.8\\
	980401	&    - &   - &    - &   - &  - &    - & 3726 & 500 & 16.4\\
	980420  & 31.5 &   5 &  9.4 & 178 &  5 & 11.1 & 2324 & 125 & 13.5\\
	980527  & 14.7 &   5 & 13.0 & 208 & 25 & 14.2 &    - &   - & -\\
	980627  &    - &   - &    - & 180 &  5 & 12.2 &  602 &  25 & 12.7\\
	981121  & 29.7 &   5 & 12.8 & 219 & 25 & 13.9 &  742 & 125 & 14.3\\
	981223  & 24.4 &   5 & 12.4 & 238 & 25 & 13.5 &    - &   - & -\\
  \end{tabular}
  \caption{Summary of limits for seven bursts responded to by ROTSE-I.
  	The time an image was taken, the exposure length, and the 
	limiting sensitivity are given.  Each time corresponds to the 
	middle of an exposure.  Multiple times are listed when the 
	sensitivity significantly improves.
	\label{tab:sevenlims}}
  \end{center} 
\end{table}

	Now that the optical signature has been seen in one case, the
natural question is whether other bursts behave like GRB990123.  One
way to address this question is to compare optical to gamma-ray
levels.  To bring all bursts onto some common footing, we correct for 
their fluence by defining:
\begin{equation}
       \umu \equiv m_V - 2.5 \log(f/f_{GRB990123}),
   \label{mudef}
\end{equation}
where $f$ is the fluence measured in BATSE channels 2 and 3 to
avoid systematics due to problems in spectral fitting the other
channels (\cite{mbriggs}).

	Several issues of optical extinction arise in our comparison.  
First, galactic
extinction varies significantly over the IPN arcs preventing us from
quoting an accurate value in most of the bursts we analyzed.  We note,
however, that for almost all of these bursts it is much less than 1
magnitude at their most probable location.  Since GRB990123 has a
similar low value (= 0.04), its effect on our comparison should be 
minimal.  The one exception is GRB980420 which may have over 2
magnitudes of absorption.  We ignore the effects of
extinction at the source because there is no measure of it in
most of these cases, aside from GRB980329a and GRB990123.  A statistical
argument has been made, however, that most GRBs are not heavily
obscured at the source (\cite{frail_obscure}).

\begin{figure}
\begin{center}
\vspace{2.0pc}
\hspace{-0.5in}
\epsfig{figure=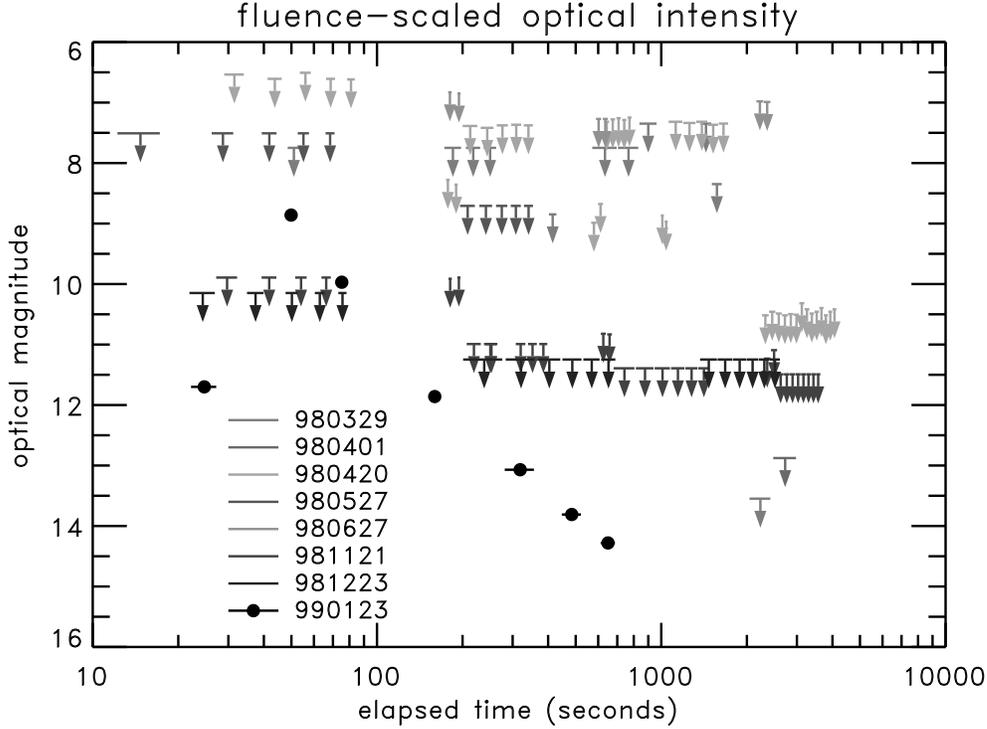,height=4.0in}
  \caption{Rescaled limits, $\umu$, for seven GRB data sets vs. $t_+$.
	Relative to GRB990123, significantly higher optical emission would
	have been seen during the first few minutes.  In two cases, the
	optical emission is at least one magnitude fainter than GRB990123.
    \label{fig:GRB_lim}}
    \end{center}
\end{figure} 

     The limits are replotted as $\umu$ vs. $t_+$ in Figure 
\ref{fig:GRB_lim}.  If the scaled optical emission was much higher 
than in GRB990123 around $t_+ =$ 50 seconds or around $t_+ =$ 300 
seconds, we would have seen it.  In several dimmer bursts, however, we 
observed no such behavior.  While we cannot rule out beaming, there is no 
evidence for it in this analysis.

	More importantly, in two cases, GRB981121 and GRB981223, the 
optical emission around one minute is at least 1 magnitude fainter
than in GRB990123.  The main implication of this analysis is that
there does not
appear to be a strong correlation of optical flux with gamma-ray fluence.
The inherent dispersion to any actual correlation must be larger than
one magnitude to explain the results from GRB981121 and GRB981223, in
particular.  A significant correlation, however, is expected in models of
reverse-shock development (\cite{sari99}).

	There are several caveats and cautions to this analysis.  The
most important is that there is a large diversity of GRB
behavior.  Our results are based on a handful of events, and aside 
from GRB980527, we are only looking at long bursts.  Another 
limitation arises from the IPN source of final positions, which 
discriminates against dim bursts.  Therefore, our results may not be
representative of GRBs in general, and more GRB triggers and further
analysis are necessary.

\section{Conclusions} 

	We have observed a prompt optical burst during the gamma-ray
emission of GRB990123.  It is as violent as the burst of gamma-rays, 
but it displays a different temporal behavior.  This difference is
consistent with the expected signature of a reverse shock from the 
explosion.

	Preliminary studies of seven other bursts reveal several
further points about GRBs.  No optical counterparts were identified,
and from this we can conclude that bright optical bursts (ie. 
$m_V \sim 10$) are uncommon.  When using fluence as an estimator of 
total energy output, no bursts with optical flux much greater than 
GRB990123 have been observed.  In two cases, the scaled optical 
emission around 1 minute is at least 1 magnitude dimmer than for 
GRB990123.  While not conclusive, the non-detection of another 
optical burst suggests that there is not a strong correlation between 
gamma-ray and optical emission.

\begin{acknowledgments}
ROTSE is supported by NASA under $SR\&T$ grant NAG5-5101, the NSF
under grants AST-9703282 and AST-9970818, the Research 
Corporation, the University of Michigan, and the 
Planetary Society.  Work performed at LANL is supported by the DOE under
contract W-7405-ENG-36.  Work performed at LLNL is supported by the DOE
under contract W-7405-ENG-48.
\end{acknowledgments}

\end{document}